\documentclass[aps,prb,preprint,showkeys]{revtex4-2}

\usepackage{amsmath}
\usepackage{graphicx, amssymb}
\usepackage{xcolor}
\usepackage[dvips]{epsfig}
\usepackage{lineno}
\begin{document} 

\title{Kinetically Decoupled Electrical and Structural Phase Transitions in VO$_2$}

\author{S. R. Sahu$^{1}$, S. S. Majid$^{2,3}$, A. Ahad$^{4}$, A. Tripathy$^{1}$, K. Dey$^{1}$, S. Pal$^{1}$, B. K. De$^{1}$, Wen-Pin Hsieh$^{2}$, R. Rawat$^{1}$, V. G. Sathe$^{1}$, , D. K. Shukla$^{1, *}$} 
\affiliation{$^1$ UGC-DAE Consortium for Scientific Research, Indore-452001, India\\
$^2$ Institute of Earth Sciences, Academia Sinica, Nankang, Taipei-11529, Taiwan\\
$^3$ Department of Physics, National Institute of Technology Hazratbal Srinagar J \& K-190006, India\\
$^4$ New chemistry unit (NCU), Jawaharlal Nehru Center for advanced Scientific Research Banglore-560064, India\\
}
\date{\today}
\begin{abstract}
Vanadium dioxide (VO$_2$) has drawn significant attention for its near room temperature insulator to metal transition and associated structural phase transition. The underlying Physics behind the temperature induced insulator to metal and concomitant structural phase transition in VO$_2$ is yet to be fully understood. We have investigated the kinetics of the above phase transition behaviors of VO$_2$ with the help of resistivity measurements and Raman spectroscopy. Resistance thermal hysteresis scaling and relaxation measurements across the temperature induced insulator to metal transition reveal the unusual behaviour of this first-order phase transition, whereas Raman relaxation measurements show that the temperature induced structural phase transition in VO$_2$ follows usual behaviour and is consistent with mean field prediction. At higher temperature sweeping rates decoupling of insulator to metal transition and structural phase transition have been confirmed. The observed anomalous first order phase transition behavior in VO$_2$ is attributed to the unconventional quasi particle dynamics, i.e. significantly lowered electronic thermal conductivity across insulator to metal transition, which is confirmed by ultrafast optical pump-probe time domain thermoreflectance measurements. 
\end{abstract}
\maketitle
\textbf{Introduction}\\
VO$_2$ undergoes a first-order insulator to metal transition (IMT) at a critical temperature (T$_c$ $\approx$ 340 K) at ambient pressure which is accompanied by structural changes, $i.e.$ transits from monoclinic insulating phase (M1: $P2_{1}/c$) into rutile metallic phase (R: $P4_{2}/mnm$) \cite{laverock2014direct,morrison2014photoinduced,qazilbash2007mott,nag2012non}. Nature of this phase transition has been discussed controversially in literature, whether it is lattice distortion driven Peierls type or electron correlation driven Mott type \cite{majid2018insulator}. Complex coupling of insulator to metal transition and structural phase transition in VO$_2$ has motivated researchers to investigate the coupling/decoupling of above transitions via external stimuli such as light \cite{wegkamp2014instantaneous,morrison2014photoinduced}, strain \cite{muraoka2014persistent,aetukuri2013control} and charge injection \cite{tao2012decoupling,nakano2012collective}. Moreover, near room temperature insulator to metal transition of VO$_2$ is extremely interesting for various important applications such as in Mott memory, Mott FET and neuromorphic devices etc \cite{zhou2015mott,ruzmetov2010three,kim2004mechanism}. Therefore, better insights about the first-order phase transition behavior of VO$_2$ is pivotal from both fundamental Physics and application point.  

 The first-order phase transition (FOPT) is not an instantaneous phenomenon but it is a dynamic process, and the nucleation and growth have their own kinetics \cite{chaddah2017first}. FOPT commonly occurs under heating or cooling cycles and exhibits hysteresis across the transition, is accompanied by sharp discontinuities in physical properties such as resistance. In FOPT, conversion from phase-1 (parent phase) to phase-2 (daughter phase) will proceed over a finite time even after the control parameter (T, in the case of temperature induced FOPT) has reached its value (\textit{T$_c$}, defined by average of the transition temperature of the heating and cooling cycles), this leads to a phase coexistence \cite{chaddah2017first}. The phenomenon of phase coexistence is broadly observed across a large number of condensed matter systems, such as high-Tc superconducting cuprates \cite{hanaguri2004checkerboard,vershinin2004local}, the colossal magneto resistive manganites \cite{zhang2002direct,lai2010mesoscopic,dho2003strain}, and also in vanadium oxides \cite{qazilbash2007mott,liu2013anisotropic,frenzel2009inhomogeneous,liu2014symmetry,mcleod2017nanotextured}. 

 During FOPT, formation of nuclei of daughter phase within the parent phase is known as the nucleation phenomenon and subsequent diffusion is the mechanism that guides and determines the phase transformation kinetics is governed by mean field predictions. Earlier reports suggest that thermal scaling of hysteresis and relaxation measurements are key experiments to test the mean field predictions during FOPT \cite{bar2018kinetic,chattopadhyay2003metastability,banerjee2011history,kushwaha2009low}. As per mean field prediction critical like slowing down of phase transformation would be observed around the transition if the system approaches a genuine bifurcation point. And a systematic delay is observed in the onset temperature that is dependent on the temperature sweeping rate (TSR) \cite{bar2018kinetic,levy2002novel,liu2016driving,perez2004driving}. Such systematic delay in the onset of phase switching is predicted and observed in various physical systems in a definite scaling form \cite{levy2002novel,liu2016driving}. Rising area of the hysteresis loop with increasing driving force sweeping rate has been theoretically predicted as well as experimentally observed in V$_2$O$_3$ \cite{zhong2017renormalization,zhong2005theory,bar2018kinetic}. 

\begin{figure}[hbt]
\centering
\includegraphics[width=\linewidth]{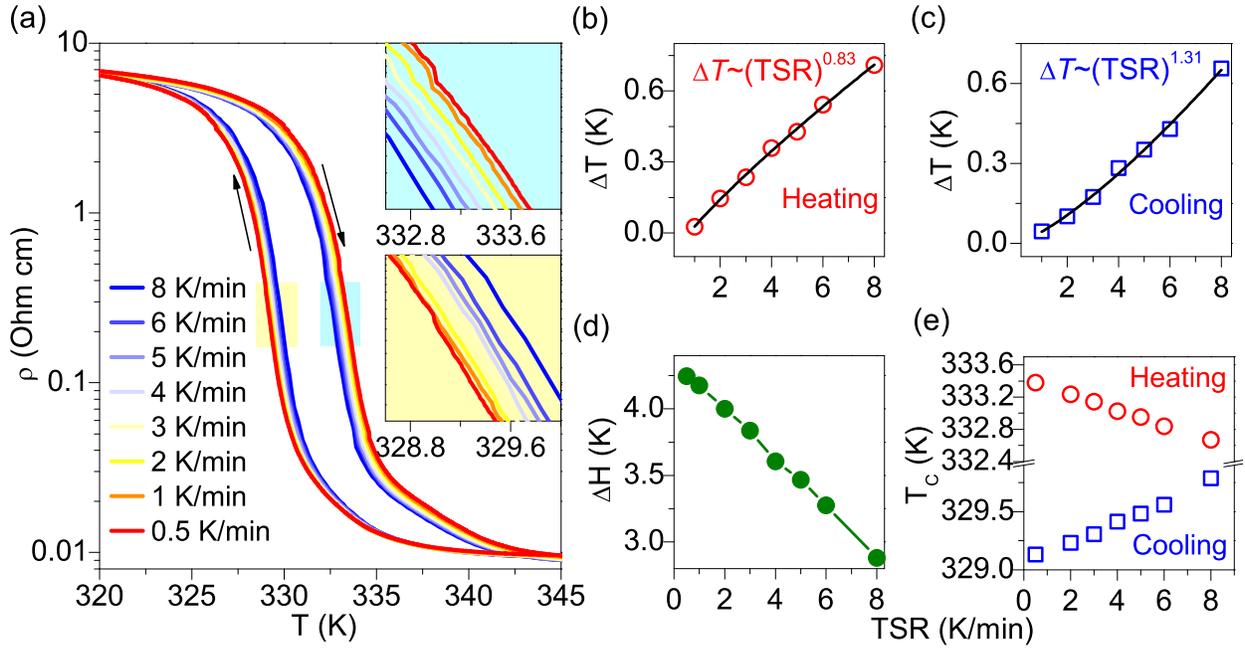}
\caption{Thermal hysteresis scaling (a) Temperature sweeping rate dependent resistivity measurements on VO$_{2}$, arrows indicate heating and cooling cycles. Insets show enlarged views of heating and cooling cycles data across T$_c$ in respective cycles. (b) and (c) show fittings of shifts in transition temperature ($\Delta T$ =$\mid {T_{c}-T_{o}}\mid$) with varying TSR in heating and cooling cycles, respectively. Hysteresis width ($\Delta H$) and transition temperature (\textit{T$_{c}$}) variations as a function of TSR is presented in (d) and (e), respectively (see methods for detail).}
\label{F1}
\end{figure}

\begin{figure}[hbt]
\centering
\includegraphics[width=\linewidth]{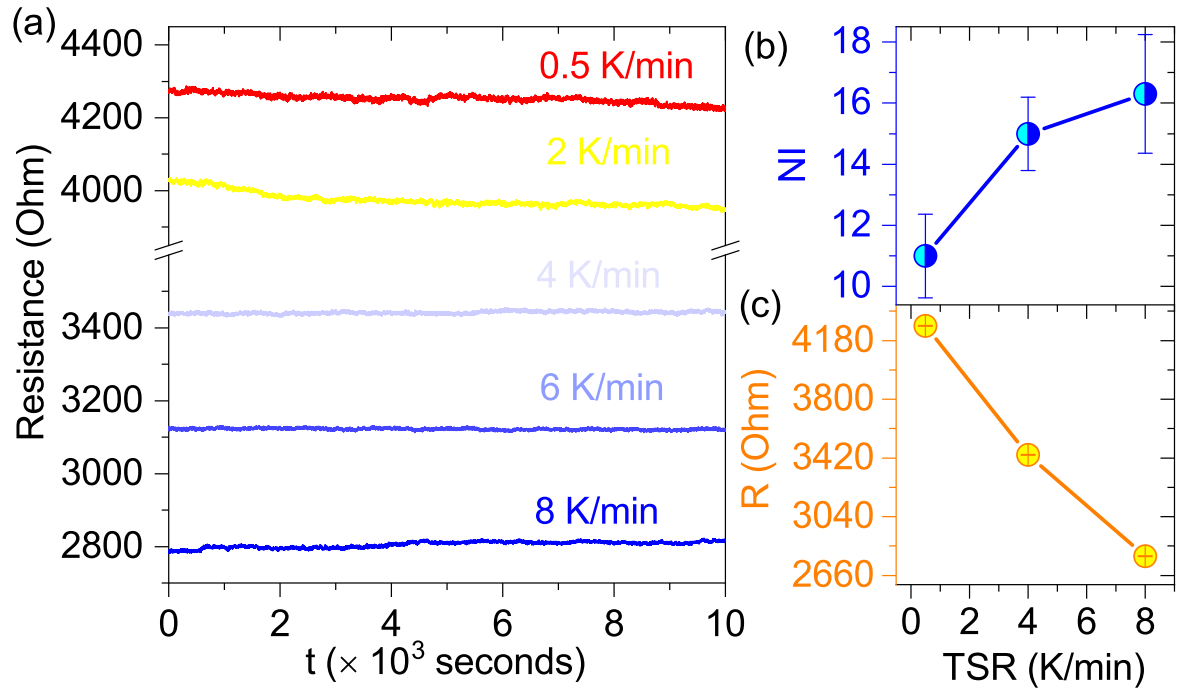}
\caption{(a) Resistance relaxation measurements performed at 332 K (\textit{T$_{m}$}), for different TSR. (b) shows the strengthening of normalized Raman intensities (NI) and (c) deceasing resistance values with increasing TSR at \textit{T$_{m}$} i.e. decreased insulating character but increased monoclinic phase with TSR.}
\label{F2}
\end{figure}

\begin{figure}[hbt]
\centering
\includegraphics[width=\linewidth]{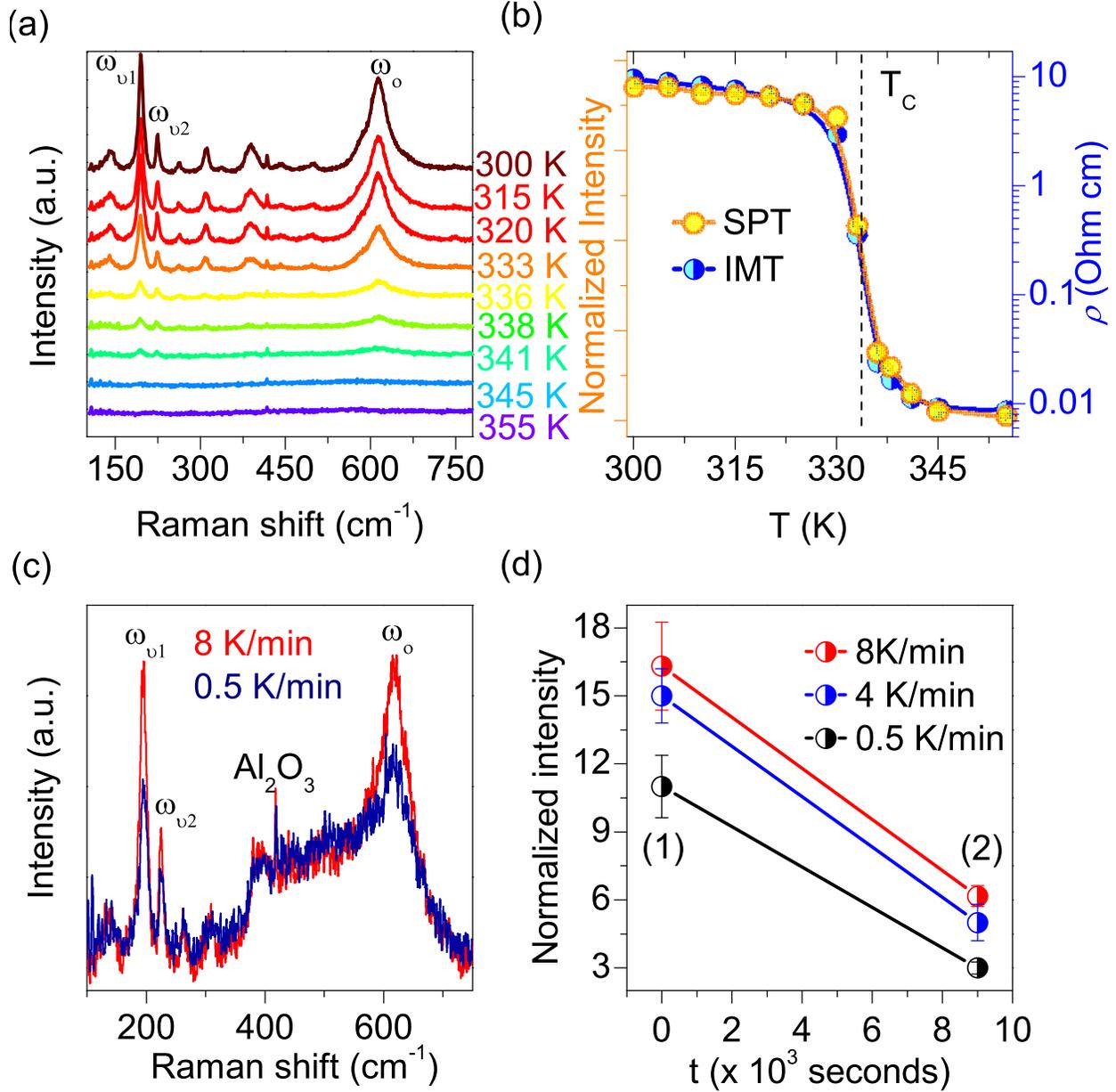}
\caption{(a) Temperature dependent Raman spectra where disappearance of Raman modes with increasing temperature confirms the structural transition from monoclinic phase to rutile phase. (b) Insulator to metal transition from temperature dependent resistivity data, and temperature induced monoclinic to rutile structural phase transition, both measured at 2 K/min. Dashed vertical line denotes the \textit{T$_{c}$} (c) Raman spectra collected immediately at \textit{T$_{m}$} after reaching to this temperature with slowest (0.5 K/min) and fastest (8 K/min) TSR. (d) Normalized intensity vs time, where data points (1) represent the Raman intensity at t = 0 (measured immediately after reaching at 332 K) and data point (2) represent the Raman intensity after t = 9000 seconds for different TSR.} 
\label{F3}
\end{figure}

\begin{figure}[t]
\centering
\includegraphics[width=\linewidth]{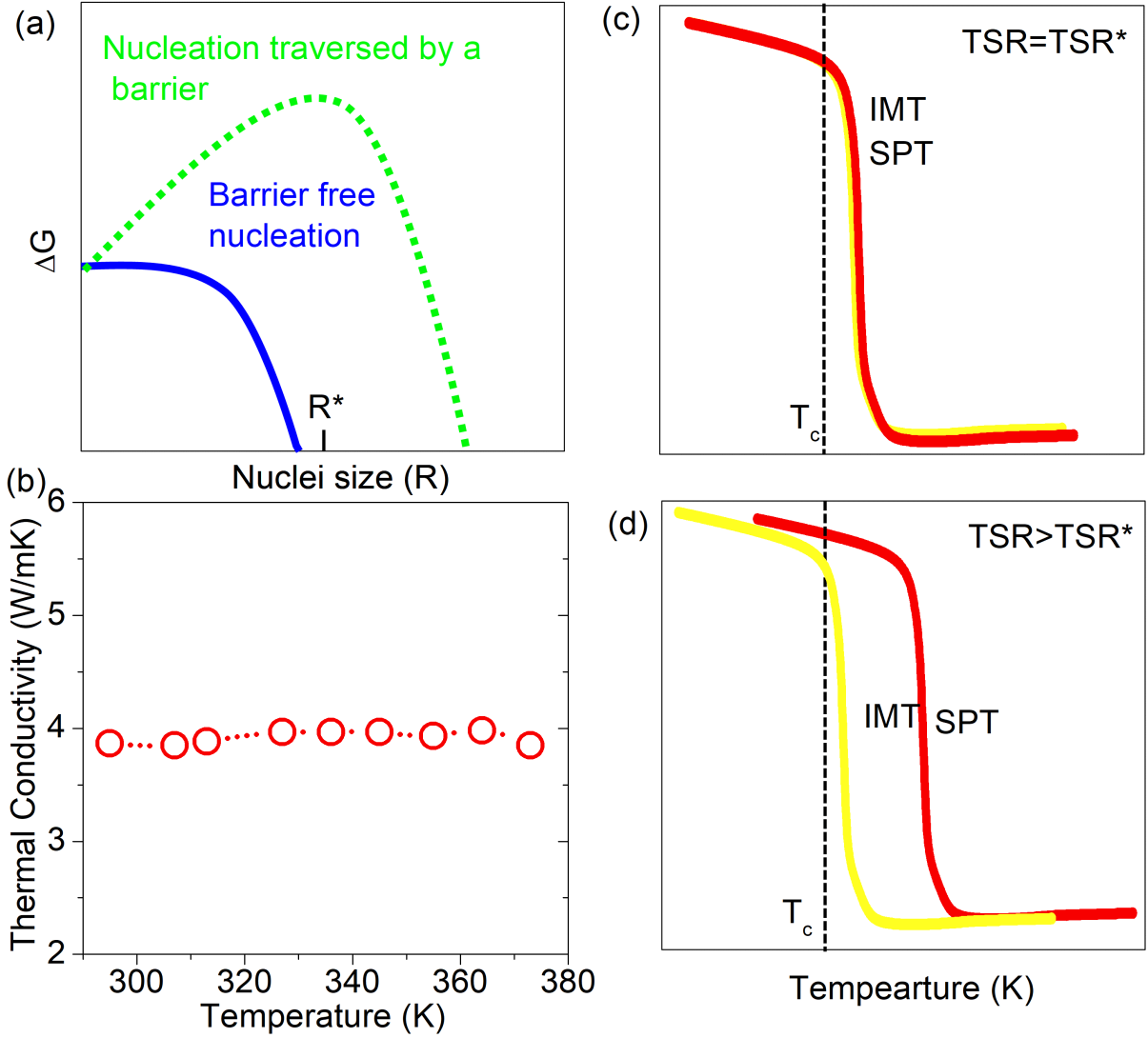}
\caption{(a) shows the processes of nucleation traversed through the barrier and barrier-free paths, during FOPT. (b) temperature dependent thermal conductivity data of VO$_{2}$ thin film. (c) shows that transition temperature is similar for TSR = TSR$^{\ast}$, where TSR$^{\ast}$ is the optimum temperature sweeping rate until when IMT and SPT are coupled. (d) is the case for TSR $>$ TSR$^{\ast}$ i.e. when IMT leads by SPT because SPT is highly time dependent process. \textit{T$_{c}$} represents the transition temperatures, while IMT and SPT curves (in (c) and (d)) are representing data of heating cycles only.}
\label{F4}
\end{figure}

 Here, we have experimentally studied the kinetics of thermally induced FOPT in VO$_2$. Minor hysteresis loop (MHL) in resistivity measurement confirmed the presence of phase-coexistence during phase transition in VO$_2$. The thermal hysteresis in resistivity measurements have been found to shrink with increasing TSR which is in contrast to the conventional FOPT. 
Observed unusual thermal hysteresis scaling behavior with TSR during insulator to metal transition may be the consequences of independent diffusion of charge and heat due to unconventional quasiparticle dynamics in VO$_2$. To substantiate the claim of unconventional quasi particle dynamics we have also performed the temperature dependent thermal conductivity measurements which clearly indicates the significantly lower electronic thermal conductivity across insulator to metal transition in the studied VO$_2$ thin film.\\\\
\textbf{Results\\}
Firstly, we have evaluated the first-order characters of the phase transition in the pulsed laser deposition grown VO$_2$ thin film. Resistivity measurement showed several orders of change in resistance magnitude and hysteresis across transition which confirmed FOPT character (see Fig. S1(c)). Next, we performed the temperature sweeping rate dependent measurements to test the mean field predictions. Resistivity relaxation measurement in the phase coexistence region has been used to track the time evolution of grown metallic phase and the Raman spectroscopy relaxation measurement is used to track the time evolution of the rutile structure. The results of above discussed experimental protocols have helped us to understand the enigma of observed anomalous FOPT.\\\\
\textbf{\textit{(i) Thermal hysteresis scaling\\}}
The dynamic hysteresis loop in Fig.\ref{F1}(a) shows the temperature dependent resistivity of the VO$_2$ thin film in the heating and cooling cycles for different TSR, ranging between 0.5 K/min to 8 K/min. Zoomed view of hysteresis loops in heating and cooling cycles are shown in insets. The shape of hysteresis loops is sensitive to the phase transformation rate \cite{ortin1992preisach}. We observed a systematic shrinking of the hysteresis width with increasing TSR, which is unusual for FOPT. A standard way to depict the dynamic shift is to plot the rate dependence of dynamical renormalization shift $\Delta T(R)$. We have plotted the variation of the dynamical renormalization shift of the transition temperature with TSR for heating (Fig.\ref{F1}(b)) and cooling (Fig.\ref{F1}(c)) cycles. Dynamical renormalization shift, $\Delta T$ is defined by $\Delta T = \mid {T_{hc}-T_{o}} \mid$ where $T_{hc}$ is the transition temperature of the corresponding resistivity curves for heating and cooling cycles performed at a particular TSR and $T_o$ is the transition temperature of quasi-static resistivity curve for heating and cooling cycles. Slowest rate here i.e. 0.5 K/min has been considered as a quasi-static case for calculations and obtained $T_o$ are 333.38 K and 329.13 K for heating and cooling cycles, respectively. The change in the area of the hysteresis loop (or equivalently, the shift in the transition point) can be scaled with sweeping rate (rate of change of temperature T as a power law) \cite{zhong2005theory,jung1990scaling,zheng1998thermal}. The $\Delta T$ vs. TSR data has been successfully fitted by the expression $\Delta T$ $\sim$ $(TSR)^\gamma$ and the values of gamma for the heating and cooling cycles have been obtained by fitting as 0.83 $\pm$ 0.07 and 1.31 $\pm$ 0.10 (see the Figs. \ref{F1} (b, c)). Variation of the hysteresis width ($\Delta H$) with varying TSR is shown in the Fig.\ref{F1}(d). Variation of the transition temperatures with TSR in heating and cooling cycles are shown in the (Fig.\ref{F1}(e)). Although these results show the strong TSR dependence of the IMT, the trend is opposite to the usual FOPT character \cite{bar2018kinetic,kushwaha2009low}.\\\\
\textbf{\textit{(ii) Resistivity relaxation measurements\\}}
To monitor the time evolution of metallic phase we have collected the resistance data at constant temperature \textit{T$_m$} $\sim332$ K (marked by dashed line in the inset of Fig.\ref{SF1} (c)) for around $\sim10000$ seconds for each TSR. For the relaxation measurement one needs a temperature in phase coexistence region where nucleation of the metallic phase has been started. Therefore,  \textit{T$_m$} is chosen in the middle of phase transition \cite{chattopadhyay2003metastability,roy2013first}. The novel protocol we have used for the relaxation measurement is as follows. Each measurement starts from the same initial temperature in insulating phase (i.e. 280 K) with different TSRs (0.5-8 K/min). As soon as \textit{T$_m$} (measurement temperature) is approached the resistance is measured for $\sim10000$ seconds. After each time dependent resistance measurement sample is cooled again to initial temperature. It is clearly observed that the trend of the relaxation data is not as that predicted by usual FOPT (Fig. \ref{F2}). For each TSR, resistance value has been observed to be almost constant all through the time evolution which is unusual. According to the usual FOPT behavior, in the transition region if one stops the transition driving force (which is temperature here) and leaves it for time evolution then it always has (whatever be the TSR) tendency to go to the daughter phase \cite{chattopadhyay2003metastability}.\\\\
\textbf{\textit{(iii) Raman relaxation measurements\\}}
Raman spectroscopy has been proven to be an invaluable tool to study the monoclinic to rutile structural transformation in VO$_2$ \cite{majid2020role,majid2018insulator}. According to group theory, the $\textit{M1}$ phase of VO$_2$ has 18 Raman active modes (9Ag and 9Bg). Out of which 12 modes at $\sim 138$, 193, 223, 261, 308, 338, 387, 439, 442, 498, 583 and 613 $cm^{-1}$ \cite{majid2020role} are observed in the VO$_2$ thin film and Raman mode at $\sim 417$ $cm^{-1}$ belongs to $Al_{2}O_{3}$ substrate (see Fig.\ref{F3}(a)) \cite{lee2016epitaxial}. The three modes labeled as $\omega_{\nu 1}$ ($\sim 193$ $cm^{-1}$) and $\omega_{\nu 2}$ ($\sim 223$ $cm^{-1}$) which belong to the the V-V vibrations \cite{marini2008optical} and $\omega_{o}$ ($\sim 615$ $cm^{-1}$) which represents V-O-V vibration (see Fig. \ref{SF1}(b)) has been treated as key modes in phase transition region studies of VO$_2$ \cite{lee2016epitaxial,majid2018insulator}. Disappearance of Raman modes with increasing temperature signifies the monoclinic to rutile structural phase transition (SPT). At slower temperature sweeping rate, temperature induced IMT and SPT are found to occur simultaneously, which is seen in the plot of temperature dependent resistivity and the normalized intensity of the Raman mode ($\omega_{\nu 1}$) of the VO$_2$ thin film, both measured in the heating cycle for 2K/min TSR (see Fig.\ref{F3}(b)). To monitor the time evolution of rutile metallic phase, Raman spectroscopy relaxation measurement is performed using the similar protocol as used for the resistivity relaxation measurements, shown in the Fig. \ref{F3}(d). For each TSR a Raman spectrum is measured immediately when \textit{T$_{m}$} (332 K) is approached (Fig.\ref{F3}(c)).  Immediately measured Raman spectra are shown in Fig.\ref{F3}(c) for the lowest TSR (0.5 K/min) and for highest TSR (8 K/min). It is clearly seen that Raman modes are stronger for 8 K/min TSR compared to that of 0.5 K/min (intensities are normalized by Al$_2$O$_3$ mode), which indicates larger monoclinic phase fraction for 8 K/min. This result is  incontrast to the resistivity data where resistance is found consistently decreasing with increasing TSR. To compare the relaxation data of Raman spectroscopy with the resistivity relaxation data, we have fitted the most intense Raman mode ($\omega_{\nu 1}$ = $\sim 193$ cm$^{-1}$) with Lorentz function and plotted the area as a function of time for different TSR (Fig.\ref{F3}(d)). Significantly decreased intensity of the Raman mode at \textit{T$_{m}$} with time indicates the growth of the rutile phase for each TSR (see Fig.\ref{F3}(d)), which is as per expectations of usual FOPT behavior. Fig.\ref{F2}(b) shows the normalized Raman intensity and Fig.\ref{F2}(c) shows the trend of resistance at \textit{T$_{m}$} (at t = 0 second) with varying TSR which clearly indicate the decoupling character of IMT and SPT with TSR.\\\\
\textbf{\textit{(iv) Thermal conductivity measurement\\}}
Our principal experimental protocol is based on temperature sweeping rate, therefore, the thermal conductivity of the material is of paramount importance in governing the kinetics of phase transition. Therefore, thermal conductivity measurements have been performed across phase transition region in VO$_2$ thin film. Thermal conductivity ($\Lambda$) has been found to be almost constant at $\sim 4$ Wm$^{-1}K^{-1}$ in the measured temperature region (see Fig.\ref{F4}(c)). These results are consistent with the earlier report \cite{lee2017anomalously}. On the phase transition into the metallic phase an increase in the thermal conductivity is expected. However, we have not observed any such rising thermal conductivity behavior with onset of metallicity, which clearly indicates the low electronic thermal conductivity of VO$_2$ in metallic phase \cite{lee2017anomalously}. Lee $et~al.$ who also measured the thermal conductivity in VO$_2$ across insulator to metal transition have confirmed that such low electronic thermal conductivity in metallic phase is a signature of the unconventional quasi particle dynamics, where heat and charge diffuse independently \cite{lee2017anomalously}.\\\\   
\textbf{Discussions\\}
 Nucleation and growth mechanism during phase transformations in solids has been a subject of great interest \cite{christian2002theory} but it is quite difficult to get a physical parameter from bulk measurements which can be directly related to the phase evolution. Generally, hysteresis accompanied with the transition is considered as an appropriate tool to determine the first order nature of the transition \cite{white2013long}. Therefore, hysteresis observed in our resistivity measurements is utilized to characterize the first order phase transition. Moreover, phase coexistence which is a typical signature of FOPT is also observed in resistivity, as seen in the MHL measurements of the VO$_2$ thin film (see Fig. S1(d)). 

The explicit character of FOPT is that with rising TSR hysteresis should broaden \cite{bar2018kinetic}. We have observed the shrinking of hysteresis with increasing TSR (Fig.\ref{F1}(a)). This is in stark contrast to FOPT character and possible only when by some means number of nucleation centers get enhanced by rising TSR \cite{suh2004semiconductor}. However, the situation for the SPT probed by Raman spectroscopy is found consistent with FOPT behavior. We have observed the less intense Raman modes for low TSR (0.5 K/min) compared to that of high TSR (8 K/min), at the same temperature (\textit{T$_{m}$}) (Fig.\ref{F3}(c)). The observed influence of TSR on Raman data clearly hints towards broadening of the SPT hysteresis with increasing TSR which is usual. In order to clearly visualize the effect of TSR on IMT and SPT, immediately obtained values of normalized Raman intensity (Fig.\ref{F2}(b)) and  resistance (Fig.\ref{F2}(c)) (at \textit{T$_{m}$} = 332 K) are plotted as a function of TSR. These two, Raman and resistivity, clearly show opposite trend. At 332 K, by increasing TSR, resistance value decreases while normalized Raman intensity increases which signify the less insulating but more monoclinic phase fraction with TSR. The delay in SPT is obvious as it is not an instantaneous process and would require time. Our observations clearly show the decoupling of the IMT and the SPT in VO$_2$ with TSR. Decoupling of IMT and SPT in VO$_2$ has been observed earlier also by charge injection method \cite{tao2012decoupling}. 

The non-equilibrium phase fraction is related to the "degree of metastability" and for a FOPT this dictates the kinetics of the growth of the daughter phase \cite{banerjee2011history}.  So, the time taken for the relaxation of the non-equilibrium phase into the daughter phase depends on growth of the respective nuclei and available interface \cite{banerjee2011history}. Surprisingly, we have observed that resistance is almost constant for all through time evolution which is not usual and hints towards a barrier free nucleation type behavior in VO$_2$. Present observations clearly show that the nuclei which are grown during temperature sweeping does not further grow (during time relaxation) when heat flux has stopped. Moreover, at constant temperature (at 332K, see Fig.\ref{F2}(a)) decreasing values of resistance with increasing TSR clearly indicates that with increasing TSR nucleation centers have also increased. These observations of time independent behavior of resistance and enhancement of nucleation centers with increasing TSR is the consequences of barrier free nucleation. In a barrier free nucleation, as soon as barrier to nucleation diminishes the nucleation process will enhance and resultant hysteresis will shrink, as observed here. Huffman $et~al.$ also reported the IMT through barrier free path in VO$_2$ thin film \cite{huffman2018highly}. Barrier free nucleation is a kind of heterogeneous nucleation and in such case the free energy of forming the critical nucleus can be altered significantly \cite{huffman2018highly}, however, in usual FOPT, homogeneous nucleation occurs only by passing through the free energy barrier ($\Delta G$) (see Fig.\ref{F4}(a)) \cite{ahn2004strain,Papon2010,huffman2018highly}.   

 On the contrary, from Raman spectroscopy relaxation measurements it is clear that normalized intensity of Raman modes are decreasing with time (Fig.\ref{F3}(d)) at each TSR. These results are consistent with the mean field prediction, i.e. monoclinic phase is going to disappear with time at \textit{T$_{m}$}, as rutile phase must evolve. The observed difference in behavior of IMT and SPT could be resultant of independent diffusion of the heat and charge in the VO$_2$, which is supported by the temperature independent behavior of thermal conductivity in VO$_2$ (see Fig.\ref{F4}(c)). Our thermal conductivity data vindicated the lower electronic thermal conductivity in temperature induced metallic phase of VO$_2$. The lower electronic thermal conductivity in metallic phase is the clear signature of unconventional quasi particle dynamics in VO$_2$.   

 After all this deliberation we come to the conclusion that at smaller TSR, IMT accompanies with SPT but beyond a critical TSR insulator to metal transition and structural phase transition decouple. TSR dependent decoupling of IMT and SPT has been summarized in Fig.\ref{F4} (c) and (d) and is understood as follows. In VO$_2$, due to unconventional quasi particle dynamics, charge and heat flows independently in phase transition regime, so, flow of charge which is responsible for metallic character will be independent from the structural transition driven by heat flow. Therefore, by increasing the TSR, VO$_2$ becomes metallic relatively earlier but does not transform into the rutile phase because structural phase transition is highly time dependent and cannot be accelerated by rising the TSR.

 Further, in order to discard the contribution of strain (which is inevitable in thin film form) in our observations and to check the reliability of our results we have also performed the TSR dependent resistivity measurements on the polycrystalline VO$_2$ pellet and we have observed the similar results i.e. shrinking of hysteresis with TSR (see Fig. S2). For further consistency check we have also performed the similar experiments on polycrystalline V$_2$O$_3$ and found broadening of hysteresis with TSR which is in accordance to mean field prediction and similar to earlier reported data (Fig. S3) \cite{bar2018kinetic}. \\\\
\textbf{Conclusions\\}
Finally, it is observed that with temperature sweeping rate insulator to metal transition in VO$_2$ exhibits the unusual FOPT however, structural phase transition with TSR probed by Raman spectroscopy is found consistent with usual FOPT. Our results clearly show the decoupling of insulator to metal transition and structural phase transition beyond a critical temperature sweeping rate. Our TSR dependent and relaxation measurement of resistivity in phase coexistence region confirms that insulator to metal transition in VO$_2$ is following the barrier free path. The unconventional quasi particle dynamics behavior (confirmed through temperature dependent thermal conductivity measurement) in the phase transition regime is held responsible for independent diffusion of charge and heat which facilitates the decoupling of IMT and SPT. Ultimately, we come to the conclusion that unconventional quasi particle dynamics can transcend the phase transformation kinetics in VO$_2$ and mean field predictions of FOPT cannot truly predict the IMT behavior, especially at higher temperature sweep rates.\\\\
\textbf{Methods}\\
\textit{(i) Sample preparation and preliminary characterization\\}
The VO$_2$ thin film used in this study is grown on r-cut Al$_2$O$_3$ substrate using pulsed laser deposition technique. A KrF excimer laser, $\lambda$= 248 nm, repetition rate of 5Hz and pulse energy of 370 mJ, was focused onto the $V_{2}O_{5}$ target with a fluence of $\sim1.1$ $J/cm^{2}$. During deposition ultrasonically cleaned r-cut Al$_2$O$_3$ substrate was maintained at a temperature of $\sim 700$$^{\circ}$C and oxygen partial pressure was maintained at $\sim$ 10 mTorr. Prior to the gas introduction, chamber was pumped down to base pressure of 1$\times$ 10$^{-6}$ Torr. Bruker D8 X-ray diffractometer with Cu K$\alpha$ radiation was used for X-ray diffraction (XRD) measurement. Transport measurements has been performed in four point probe configuration using Keithley 2401 source meter and Keithley 2182A nanovoltmeter. Temperature was controlled using Cryocon 22C temperature controller. Raman spectroscopy is carried out in back scattering geometry using 8 mW Ar (473 nm) laser as an excitation source coupled with a Labram-HR800 micro-Raman spectrometer. Room temperature XRD Fig.S1(a) and Raman spectra Fig.S1(b) confirm the monoclinic $\textit{M1}$ phase in the VO$_2$ thin film \cite{majid2020role,majid2018insulator}. Several orders of magnitude change in resistivity across the IMT reflects the high crystalline quality of the VO$_2$ thin film, Fig.S1(c). In order to completely characterize the IMT, transition temperature (\textit{T$_{c}$}) and hysteresis width ($\Delta H$) is calculated from the Gaussian fitting of the differential curve of lnR vs. T plot. $\Delta H$ defined as the difference in the transition temperature for heating and cooling and \textit{T$_{c}$} is defined by the average of the transition temperatures of the heating and cooling cycle.\\\\
\textbf{\textit{(ii) Phase coexistence in VO$_2$ thin film, and thermal hysteresis scaling in bulk VO$_2$\\}}
The area of minor hysteresis loops (MHLs) is the convenient parameter which is related to the phase fraction and MHL can be used to verify the phase coexistence across the FOPT \cite{manekar2008nucleation}. We have drawn the MHL around the $\sim$ 321 K (Fig.\ref{SF1}(d)) which corresponds to the coexistence region of both phases, insulating and metallic. For consistency, we have also studied bulk VO$_2$. We have used polycrystalline VO$_2$ powder, procured form Sigma-Aldrich Corporation, which was pelletized for transport measurements. One can clearly see that the polycrystalline sample also exhibits the anomalous FOPT character (see Fig.\ref{SF2}). The area of the hysteresis loop is decreasing with increasing TSR. This corroborates that this is inherent property of the VO$_2$ and strain can not be held responsible for observed anomalous FOPT character.\\\\
\textbf{\textit{(iii) Thermal hysteresis scaling in bulk V$_2$O$_3$\\}}
We have also measured polycrystalline V$_2$O$_3$ powder, procured form Sigma-Aldrich Corporation. Pressed pellet was used for transport measurements. One can clearly see that the area of hysteresis loop is increasing with rising TSR, which is a typical signature of usual FOPT (see Fig.\ref{SF3}) and this result is consistent with mean field predictions. Asymmetry seen in heating and cooling cycles is because rate in cooling cycle is limited by cooling efficiency (liquid nitrogen is used for cooling and sample is in vacuum;  cooling is majorly through convection process and boiling temperature of liquid nitrogen is 77 K), therefore a systematic delay in transition temperature in cooling cycle is not observed. Such large asymmetry has not been observed in VO$_2$ because transition temperature of VO$_2$ is relatively very high than liquid nitrogen boiling temperature and due to the large temperature gradient we get such high cooling efficiency.\\\\
\textbf{\textit{(iv) Thermal conductivity measurement in VO$_2$ thin film\\}}
Thermal conductivity of the VO$_2$ thin film was measured by a well-developed ultrafast optical pump- probe Time Domain Thermoreflectance (TDTR) technique. In the TDTR measurements, the pump and the probe beams obtained by splitting of 785 nm wavelength from mode locked Ti: Sapphire laser, were focused on the VO$_2$ sample coated with $\sim 86$ nm Al film. The Al film on the VO$_2$ thin film was heated by the pump beam and the resulting in the optical reflectivity changes as a function of time were measured by the probe beam. To detect small variations in the in phase (V$_{in}$) and out of phase (V$_{out}$) components of the reflected probe beam, the measurements were synchronized with 8.7 MHz modulation frequency of the pump beam and were detected and read by fast silicon photodiode and lock-in amplifier. Detailed description of the TDTR method is reported elsewhere \cite{cahill2004analysis,hsieh2009pressure}. For the temperature dependent measurements, the sample stage was coupled with the temperature controller that measures and controls the temperature of the sample.

 Thermal conductivity of the VO$_2$ thin film was estimated by comparing the ratio -V$_{in}$/V$_{out}$ as function of delay time between pump and probe beams using a bidirectional thermal model that considers heat flowing from the heated Al film into the 70 nm VO$_2$ thin film and the sapphire substrate \cite{cahill2004analysis,oh2010thermal}. The thermal model includes several parameters like laser spot size ($\sim 7.6$ m in radius), thickness of the Al ($\sim 86$ nm, calculated by picosecond acoustics), VO$_2$ thin film and sapphire layers and the volumetric heat capacity of each layer. Thermal conductivity of Al has been taken as 200 Wm$^{-1}K^{-1}$ \cite{zheng2007high}, while temperature dependent thermal conductivity of the bare sapphire substrate had been measured separately having $\sim35$ W$m^{-1}K^{-1}$ value at room temperature (RT). For thermally VO$_2$ thin film on high thermal conductivity sapphire substrate, the sensitivity of the measurements to the variation of heat capacity is negligible \cite{oh2010thermal}. In this context we have fixed the value of specific heat capacity as 3 Jcm$^{-3}K^{-1}$ below the IMT temperature and 3.6 Jcm$^{-3}K^{-1}$ above the IMT temperature in the thermal model \cite{zheng2007high}. Thermal conductivity of the VO$_2$ thin film is the only significant unknown and free parameter to be determined. Tests of sensitivity of the thermal model to input parameters below and above the IMT temperature are shown in supporting information Fig.\ref{SF4}.\\\\

\textbf{Acknowledgments}\\ 
DKS acknowledges support from SERB New Delhi, India in the form of early career research award (ECR/2017/000712). S.S.M. acknowledges the financial support from SERB, India, in the form of the national post doctoral fellowship (NPDF) award (PDF/2021/002137/PMS). Martin Greven and Damjan Pelc are gratefully acknowledged for fruitful suggestions.\\\\
\textbf{Author contributions}\\ 
DKS conceived the idea. DKS and SRS designed the experiments and SRS developed the necessary data acquisition software and performed measurements with the help of AA, AT and KD. SSM synthesized the thin film and performed the preliminary characterization. SSM and WH performed the thermal conductivity measurements. BKD and VGS performed the Raman spectroscopy measurements. SRS, SSM, AA, SP, RR and DKS analyzed the data. DKS and SRS wrote the manuscript with the feedback from all the co-authors.\\\\
\textbf{Competing interest}\\ The Authors declare no competing financial or non-financial interest.\\\\
\textbf{Corresponding author$\ast$:} D. K. Shukla (dkshukla@csr.res.in)\\\\
\textbf{Data availability}\\ All data and materials used to generate results in the paper are available from the corresponding author upon request.\\\\

\renewcommand{\thefigure}{S\arabic{figure}}
\setcounter{figure}{0}
\begin{figure}[t]
\centering
\includegraphics[width=\linewidth]{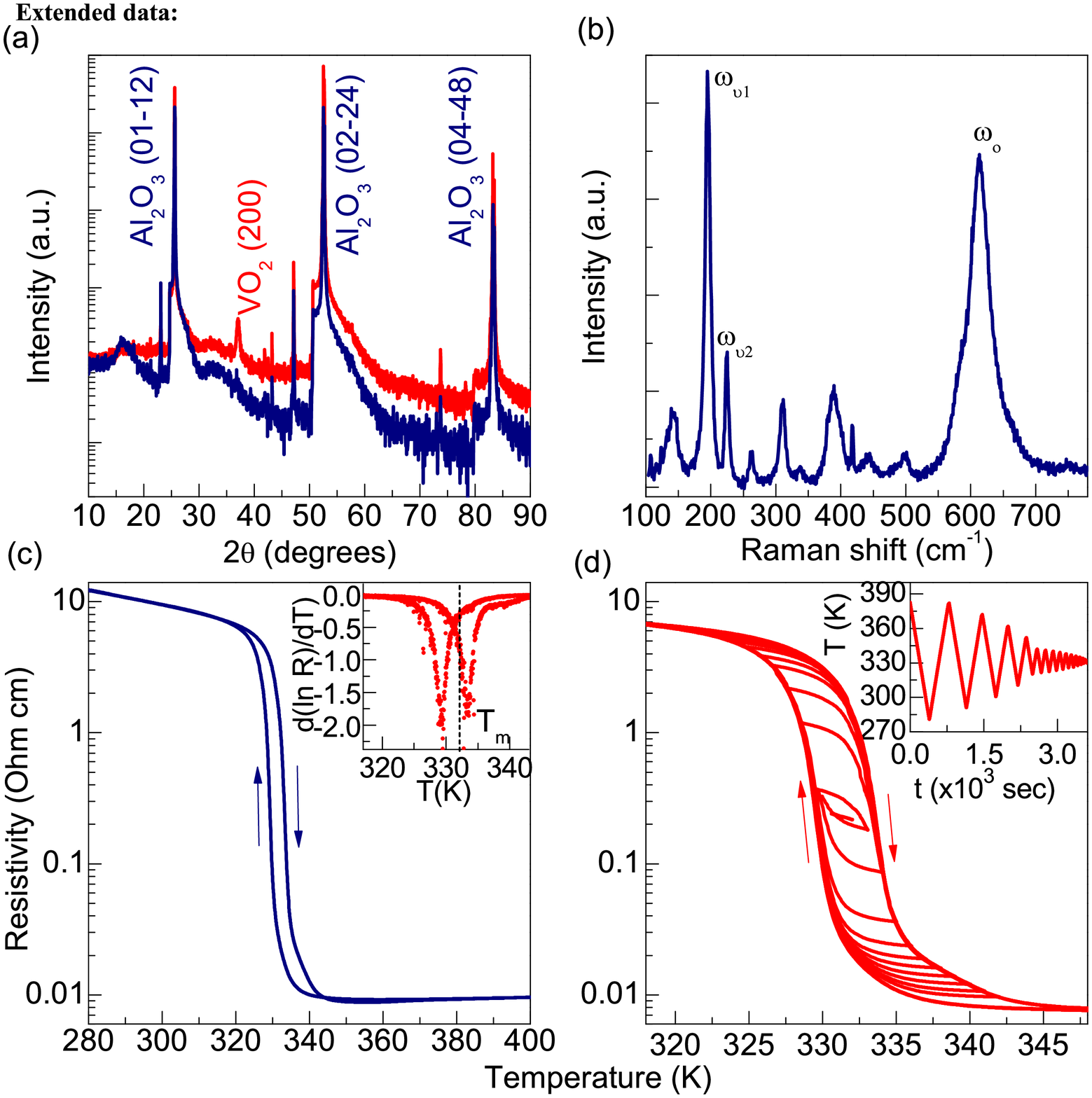}
\caption{(a) and (b) show the room temperature XRD patterns of VO$_2$ (along with that of substrate) and Raman spectra of VO$_2$ thin film grown on r-cut Al$_2$O$_3$ substrate, respectively. (c) Temperature-dependent resistivity of VO$_2$ showing the insulator to metal transition (IMT), Inset: differential curves of resistance with temperature, \textit{T$_{m}$} is marked by dashed line. (d) Multivaluedness of the sample resistance exhibiting the metastable states in the hysteresis region. Inset of Fig. 1(d) shows the temperature-time protocol used to draw the minor hysteresis loop.}
\label{SF1}
\end{figure}

\begin{figure}[t]
\centering
\includegraphics[width=\linewidth]{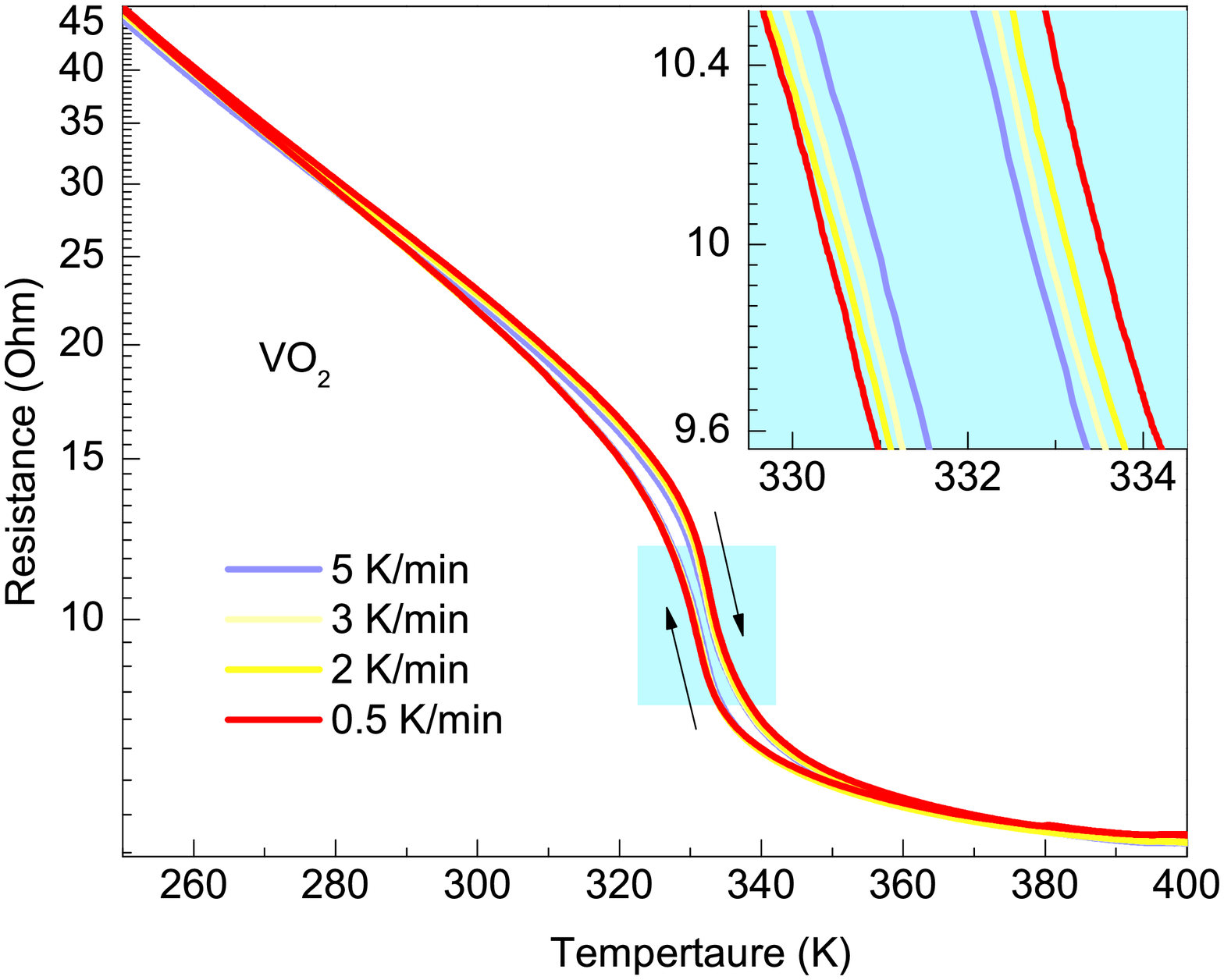}
\caption{Temperature dependent resistance measurement performed under various TSR on the bulk VO$_2$ pellet.}
\label{SF2}
\end{figure}

\begin{figure}[t]
\centering
\includegraphics[width=\linewidth]{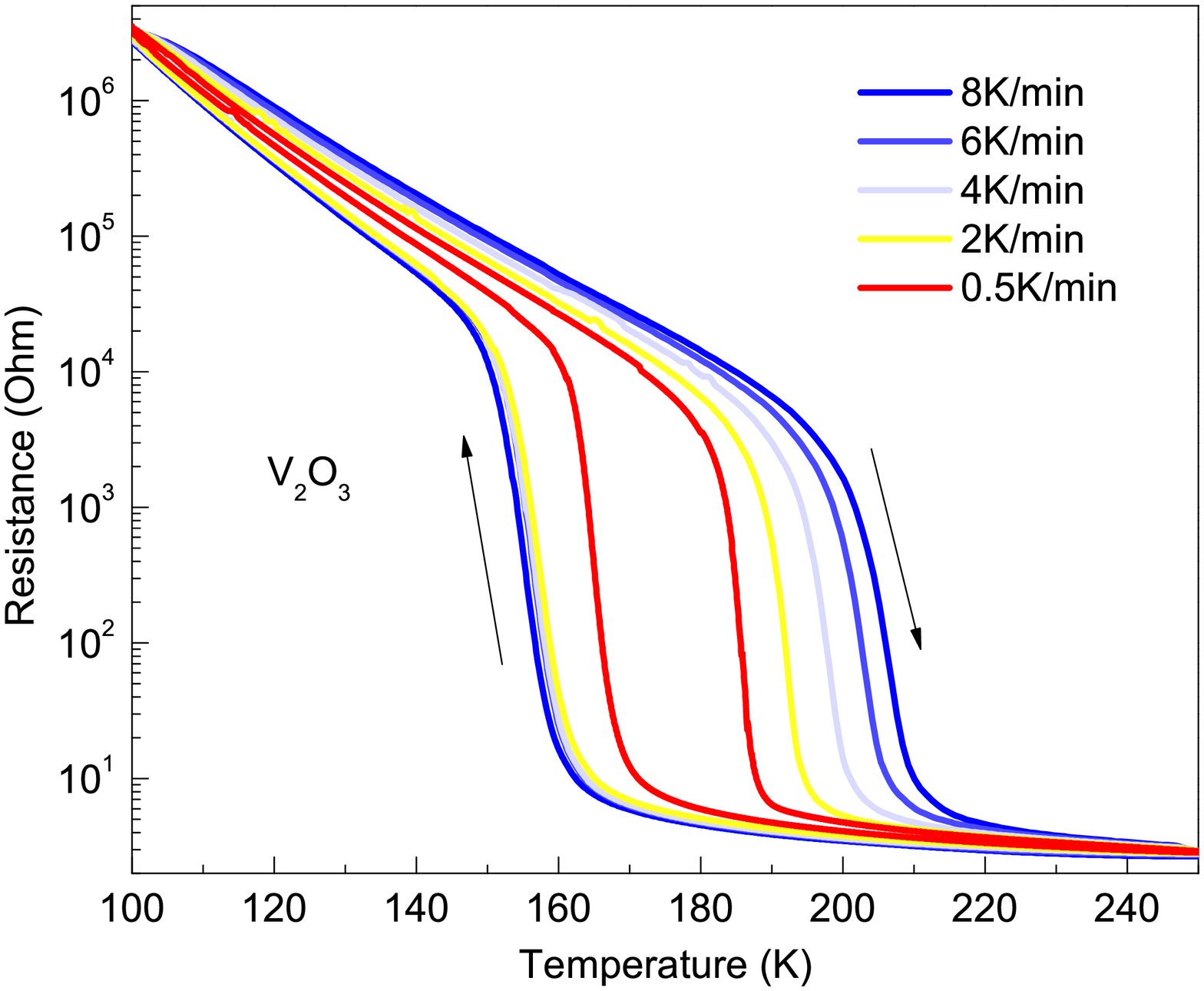}
\caption{Temperature dependent resistance measurement performed under various TSR on the bulk V$_2$O$_3$ pellet.}
\label{SF3}
\end{figure}

\begin{figure}[t]
\centering
\includegraphics[width=\linewidth]{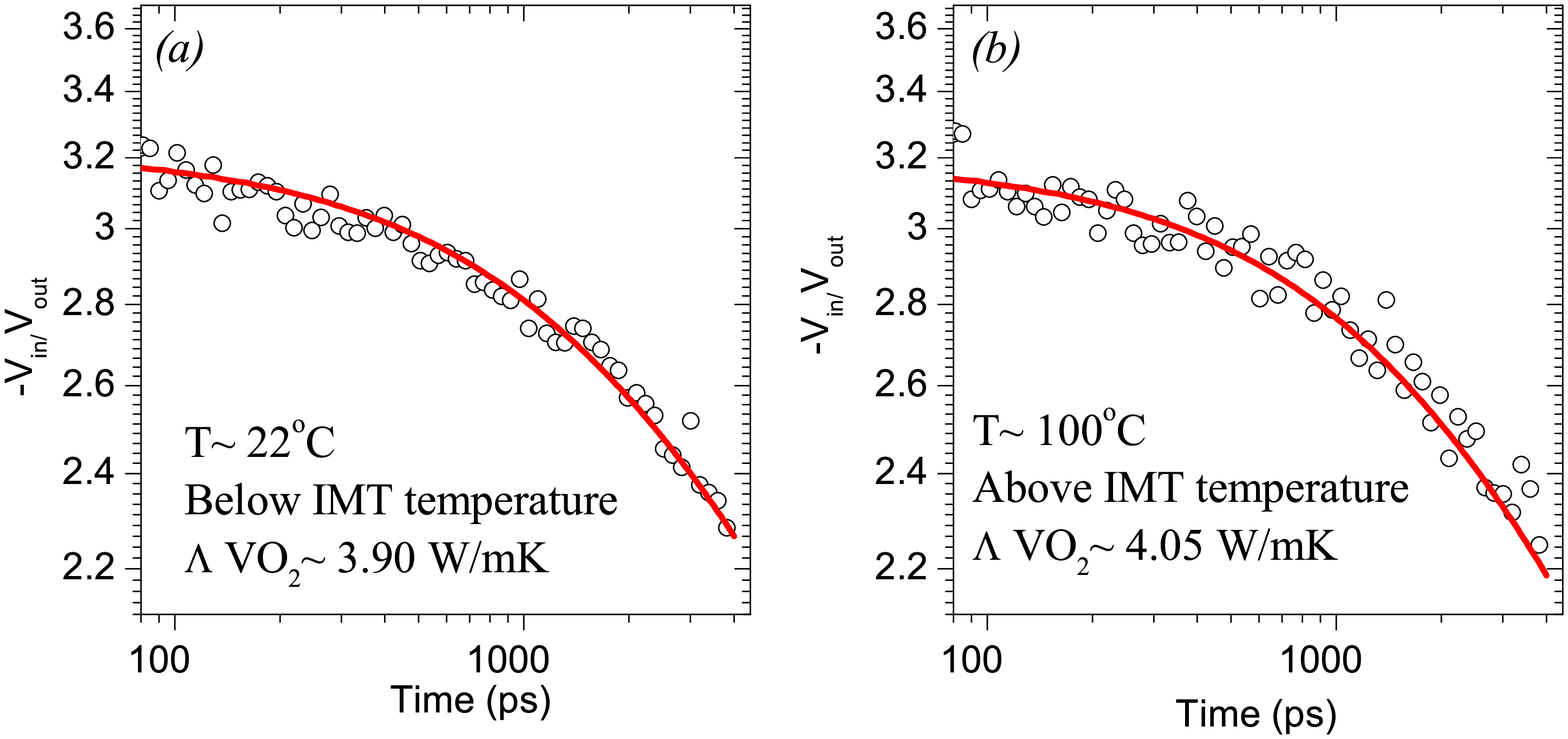}
\caption{Time domain thermoreflectance spectra of the VO$_2$ in film at (a) 295 K and (b) 373 K temperatures, fitted by the thermal model calculation (red curve). Experimental data for the ratio -V$_{in}$/V$_{out}$ as a function of delay time between pump and probe beams are shown as open circles. The best fit to the experimental data is produced with the thermal conductivity ($\Lambda$) values $\sim 3.9$ W$m^{-1}K^{-1}$ and $\sim 4.05$ W$m^{-1}K^{-1}$ of the VO$_2$ thin film at respective 295 K and 373 K temperatures.  Under the experimental conditions, the ratio -V$_{in}$/V$_{out}$ is sensitive to the thermal conductivity of the VO$_2$ thin film during delay time of few hundred picoseconds, in particular from 100 to 500 ps \cite{zheng2007high,cahill2004analysis}.}
\label{SF4}
\end{figure}

\clearpage
\bibliography{VO2FOPT}

\end{document}